\documentclass[amssymb,aps,twocolumn,floats]{revtex4}
\usepackage{color,graphicx,pstricks}
\draft
\begin{document}

\title{ Charge ordering and magnetism in quarter-filled Hubbard-Holstein model}

\author{Sanjeev Kumar$^{1,2}$ and Jeroen van den Brink$^{1,3}$}

\affiliation{
$^{1}$ Institute Lorentz for Theoretical Physics, Leiden University,
P.O. Box 9506, 2300 RA Leiden, The Netherlands \\
$^{2}$ Faculty of Science and Technology
University of Twente, P.O. Box 217, 7500 AE Enschede, The Netherlands \\
$^{3}$ Institute for Molecules and Materials, Radboud Universiteit Nijmegen,
P.O. Box 9010, 6500 GL Nijmegen, The Netherlands
}

\begin{abstract}
We study a two-dimensional Hubbard-Holstein model with phonons treated in
the adiabatic limit. A Hartree-Fock
decomposition is employed for the Hubbard term. A range of electronic densities are discussed
with special emphasis on the quarter-filling ($n=0.5$). We argue that the quarter-filled system is relevant 
for the electronic properties observed at the interface between LaAlO$_3$ and SrTiO$_3$, where 
half-electron per unit cell is transferred to the TiO$_2$ layer as a consequence of the polar 
discontinuity at the interface. 
In addition to presenting the overall phase diagrams, we identify an interesting charge-ordered 
antiferromagnetic phase for 
$n=0.5$, which was also reported recently in the ab-initio
study of the LaAlO$_3$-SrTiO$_3$ interface.

\vskip 0.2cm


\vskip 0.2cm

\today

\end{abstract}

\maketitle

\section{Introduction}

The interfaces between bulk insulating oxides such as LaAlO$_3$ (LAO) and SrTiO$_3$ (STO) have recently
become a topic of very active research \cite{hwang1,hwang2,thiel,thiel2,reyren}. 
The interest in such interfaces
was triggered by the obervation of unusually high
in-plane conductivity at the TiO$_2$-terminated interface \cite{hwang2}. Materials with
such high mobilities can find numerous applications in various fields of industry 
and can also be utilized to make new devices \cite{thiel,thiel2}.
The origin of this effect is believed
to be an electronic reconstruction caused by the polar discontinuity at the interface \cite{hwang3}.
The conceptual
idea is that in order to avoid a divergence of electrostatic potential, $0.5$ electrons per unit cell
are transferred to the TiO$_2$ layer. These electrons then behave as a quasi two-dimensional electron gas
leading to large mobilities.

Subsequently, experiments were carried out at variable oxygen
pressure and a strong dependence in the transport measurements was reported \cite{brinkman}.
In the presence of high
oxygen pressure, an insulating behavior in the resistivity was observed at low temperatures. 
This indicates that perhaps
oxygen vacancies play a crucial role in the existence of high interfacial conductivity
and the system may actually be insulating in the absence of oxygen vacancies.
In addition,
an external magnetic field has a strong effect on transport \cite{brinkman}.
Applying magnetic field leads to a
gain in the conductivity at low temperatures. An explanation for the insulating behavior of the 
resistivity was suggested to be connected to the magnetism and perhaps to the Kondo effect.

While the concept of electronic charge reconstruction or oxygen vacancies can provide a reasonable
explanation
for the observed conductivity, there is no rigerous understanding for the origin of
magnetism at the interface. Hence, the understanding of magneto-transport is also rather incomplete.
The present theoretical understanding of the interface properties has been largely based on the Density 
Functional Theory (DFT) calculations. Existence of a charge ordered state has been pridicted by these
calculations \cite{pentcheva}.
At the level of a minimal model, a charge-ordered state can be obtained within the
extended Hubbard model \cite{kopp}.
But, such a state should be non-magnetic, suggesting that there might be
an alternate mechanism active in these systems which leads to the charge ordering.
Recent LDA+U studies of the LAO/STO interface have shown that structural distortions are present
in the vicinity of the interface indicating the presence of an electron-lattice coupling. In this
study a charge-ordered antiferromagnetic state was found to be the
groundstate \cite{zhicheng}.

In this paper we study the two-dimensional Hubbard-Holstein model as
a simplest model capturing the effects of both, the electron-electron and
the electron-lattice interactions.
We find that the magnetic moments are formed as a consequence of polaron formation.
These magnetic moments are found to be antiferromagnetically correlated at finite densities.
At quarter filling, the groundstate is antiferromagnetic and charge-ordered in agreement with
the findings of recent LDA+U calculations.

\section{Model and Method }

We consider a one-band Hubbard-Holstein model on a square lattice with the
Hamiltonian:

\begin{eqnarray}
H &=& -t \sum_{\langle ij \rangle \sigma}
\left ( c^{\dagger}_{i \sigma} c^{~}_{j \sigma} + H.c. \right )
 + U \sum_i n_{i\uparrow} n_{i\downarrow} \cr
&& - \lambda \sum_i x_i (n_i - n)
+ {K \over 2} \sum_i x_i^2. ~ ~ ~ ~ ~ ~ ~ ~
\end{eqnarray}

\noindent
Here, $c^{}_{i \sigma}$ and $c^{\dagger}_{i \sigma}$ are the electron annihilation and creation 
operators with electronic spin $\sigma = \uparrow, \downarrow$. $n_i = n_{i\uparrow}+n_{i\downarrow}$ 
is the charge density operator at site i with $n_{i\sigma}=c^{\dagger}_{i \sigma} c^{~}_{i \sigma}$. 
The average electronic charge density is denoted by $n$ and
$x_i$ denote the volume contraction and expansion of the oxygen octahedra, which couples to
the variations in the charge density.
The hopping parameter $t$ is set to 1, therefore all energy scales are in units of $t$.
$U$ is the strength of on-site Hubbard repulsion, and $\lambda$ is that of electron-lattice
coupling. The lattice stiffness constant $K$ is set to 1.

In the present study the lattice distortions $x_i$ are treated in the adiabatic limit. In the absence 
of the Hubbard term, the electronic Hamiltonian is
bilinear in annihilation and creation operators, with the background potential provided by the configuration $\{x_i\}$ of
lattice distortions. The groundstate therefore, corresponds to the lattice configuration which minimizes the total energy.
In the absence of the kinetic energy term ($t=0$), the problem reduces to $N$ replicas of the
one on a single site, where $N$ is the number of lattice sites. The total energy for a single
site is given by,

\begin{eqnarray}
E &=& -\lambda x_i (\langle n_i \rangle -n) + x_i^2/2.
\end{eqnarray}

\noindent
Here and below $\langle A \rangle$ denotes expectation value of the operator $A$.
Minimization of the energy $E$ w.r.t the classical variable $x_i$ leads to 
$ x_i = \lambda (\langle n_i \rangle -n)$.
For finite $t$, however, the kinetic energy term also contributes to the total energy,

\begin{eqnarray}
E &=& -t \sum_{\langle ij \rangle \sigma}
\left \langle c^{\dagger}_{i \sigma} c^{~}_{j \sigma} + H.c. \right \rangle
 \cr
&& -\lambda \sum_i x_i (\langle n_i \rangle -n) + 1/2 \sum_i x_i^2.
\end{eqnarray}

\noindent
Note that $x_i$ is not a specified potential, but has to
be determined self-consistently with the distribution of the electronic charge density.
We compute the self-consistent lattice potential in the following scheme:
Start with an arbitrary configuration of lattice distortions $\{ x_i \} $. 
Diagonalize the Hamiltonian to generate
the eigenvalues and eigenvectors. Compute the electronic charge density $\langle n_i \rangle$.
Use the relation $ x_i = \lambda (\langle n_i \rangle -n) $ at each site to generate the new 
configuration for $\{ x_i \}$.
Repeat the process until the old and new charge densities match within given error bar.

In order to include the Hubbard term into the above self-consistent formalism, we treat the
Hubbard term within Hartree-Fock approximation.
The Hartree-Fock decomposition of the Hubbard term leads to,
\begin{eqnarray}
H_U = U \sum_i \langle n_{i\uparrow} \rangle n_{i\downarrow} + n_{i\uparrow}
\langle n_{i\downarrow} \rangle - \langle n_{i\uparrow} \rangle \langle n_{i\downarrow} \rangle.
\end{eqnarray}

\noindent
Now the self-consistency cycle requires the convergence of $\langle n_{i\uparrow} \rangle$ and
$\langle n_{i\downarrow} \rangle$
individually. The generic problem with the self-consistent methods is that they need not lead to the
minimum energy solution. Therefore, we use a variety of ordered and random initial states for the
self-consistency loop and select the converged solution with the lowest energy.

The Hubbard-Holstein model contains a variety of interesting phases and phenomena including,
superconductivity,
charge- and spin-density wave formation, phase separation, and polaron and bipolaron formation.
For this reason, the Hubbard-Holstein model has always been of interest in different contexts
\cite{pao,mac,capone,capone1,takada,clay}.
Most of the earlier studies on this model were focused at or near half-filling.
The quarter-filled
case has not been analyzed in much detail except in one-dimension \cite{hardikar}.

\section{Results}

\subsection{Dilute limit}

We begin by analyzing the case with very few electrons.
Consider the Hamiltonian Eq. (1) with a single electron. The Hubbard term is
inactive and for small $\lambda$ the groundstate wavefunction corresponds
to a Bloch wave. Since the lattice remains undistorted {\it i.e.} $x_i \equiv 0$, the
only contribution to the total energy is from the kinetic energy. For a single electron in
a 2D square lattice, the lowest eigenvalue is $-4t$, which is also equal to the total kinetic
energy. Upon increasing the value of
$\lambda$, the energy is gained via the Holstein coupling term by self-trapping of the electron into a
single polaron (SP). In mean field the trapping occures only when the energy of the SP state is
lower than $-4t$. Within a simple analysis, where we assume an ideal trapping of the
electron at a single site, Eq. (2) leads to, $E_{SP} = -\lambda^2 + \lambda^2/2 = -\lambda^2/2$. 
Hence, the critical
value of $\lambda$ required for trapping a single electron into a single polaron is given by
$\lambda_c^{SP} = 2 {\sqrt 2}$.

Now consider the case of two electrons. In addition to the possibility of trapping the electrons
as two single polarons, it is also possible to find a bipolaron (BP) solution. Infact the BP solution has
a lower energy than two SPs and the critial value of $\lambda$ required to form a BP is given by
$\lambda_c^{BP} = 2$.
The tendency to form bipolarons is clearly suppressed by the repulsive
energy cost of the Hubbard term. From a very simple analysis of the two electron case
one obtains:
$E_{free} \sim -8$, $E_{BP} \sim -2\lambda^2 + U$ and 
$E_{SP} \sim -\lambda^2 $. Looking for various energy crossings as a function of
$\lambda$ and $U$ one obtains the phase diagram shown in Fig.1(a) for the three phases
considered above. The solid lines are from the simple analysis described above and the symbols 
represent the boundary values from the self-consistent numerical calculation.
Note that this phase diagram corresponds to the case of two electrons in
an infinite lattice and therefore refers to $n \rightarrow 0$ in terms of fractional electronic filling
in the thermodynamic limit.

\begin{figure}[t]
\centerline{
\includegraphics[width=8.8cm , clip=true]{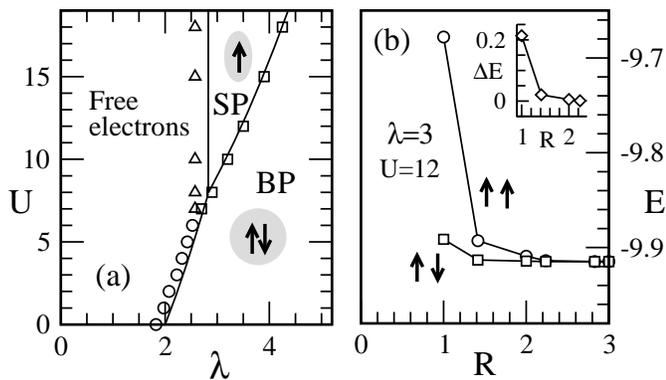}
}
\vspace{.1cm}
\caption{~(a) Phase diagram in the limit of low electron density in the parameter space of
electron-lattice coupling $\lambda$ and the Hubbard repulsion $U$.
Single polaronic and bipolaronic regimes are denoted by SP and BP, respectively.
The solid lines are from a strong-trapping analysis (see text) and the symbols
are the results of numerical calculations.
(b) Total energy as a function of the distance between two single polarons. The circles (squares)
are for the parallel (antiparallel) spins of the single polarons. Inset shows the
effective magnetic coupling between two self-trapped electrons as a function of the distance between them.
}
\end{figure}

Assuming that $U$ is large so that we are in the regime of single polaron formation, we estimate
the effective interaction between two single polarons by
calculating the total energy
as a function of the distance between them.
The energy difference between the spin-alligned and
spin-antialligned single polarons provides an estimate for effective magnetic interaction between two 
polarons. The energy variations are shown in Fig. 1(b), suggesting a repulsive and
antiferromagnetic interaction between the localized magnetic moments. The energy difference 
$\Delta E = E_{\uparrow \uparrow}-E_{\uparrow \downarrow}$ is plotted in inset in Fig. 1(b).
Positive values of
$\Delta E$ for all $R$ show that the two trapped moments prefer to be antiferromagnetic for all
distances. In fact, the strength of the interaction is almost vanishingly small for $R>2$, suggesting 
the absence of any ordered magnetic state for low densities. We will see in the following sections that
the above analysis of the dilute limit provides a very simple understanding of the phases that occur
at finite densities, in addition to clarifying the basic competing tendencies present in the 
Hubbard-Holstein model.

\subsection{Generic Electron Densities}

For analyzing the system at higher electron densities we employ the self-consistent
method described in the previous section. For the converged solution with minimum energy, we
compute the charge structure factor,

\begin{eqnarray*}
D_n({\bf q})=N^{-2}\sum_{ij} (\langle n_i \rangle -n) ( \langle n_j \rangle -n) ~
e^{-{\rm i}{\bf q}\cdot ({\bf r}_i-{\bf r}_j)},
\end{eqnarray*}
\noindent
and the spin structure factor,

\begin{eqnarray*}
D_s({\bf q})=N^{-2}\sum_{ij} \langle s_i \rangle \langle s_j \rangle ~
e^{-{\rm i}{\bf q}\cdot ({\bf r}_i-{\bf r}_j)},
\end{eqnarray*}
\noindent
with $s_i = (n_{i \uparrow} - n_{i \downarrow})/2$.
Various ordered phases are inferred from the peaks in these structure factors.
\begin{figure}[t]
\centerline{
\includegraphics[width=8.4cm , clip=true]{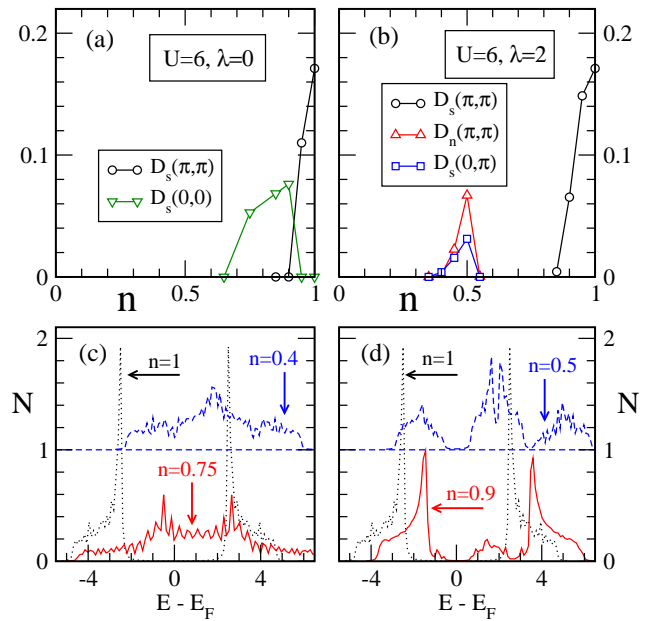}
}
\vspace{.1cm}
\caption{(Color online) ~ The charge and spin structure factors at various ${\bf q}$ as a function of
average charge density $n$ at $U=6$ for, (a) $\lambda = 0$ and (b) $\lambda = 2$.
The total density of states for selected $n$ are shown for, (c) $\lambda = 0$ and
(d) $\lambda = 2$. The dotted (blue) curves in (c) and (d) are shifted along the y-axis for clarity.
}
\end{figure}
Fig. 2(a) shows the spin structure factor at ${\bf q} = (\pi,\pi)$, which is a measure of antiferromagnetic
correlations, and at ${\bf q} = (0,0)$, which is indicative of a ferromagnetic behavior.
At $\lambda = 0$ The system is antiferromagnetic (AFM) at and near $n=1$, becomes ferromagnetic (FM) for
$0.7 < n < 0.9$, and  eventually becomes paramagnetic (PM). The antiferromagnetism at half-filling 
arises as a consequence of the nesting feature of the Fermi surface.
The ferromagnetism at intermediate densities can be understood within a Stoner picture which
suggests that the repulsive cost coming from the Hubbard term can be reduced by a relative shift
of the spin-up  and spin-down bands.
At $\lambda = 2$, the antiferromagnetic regime near $n=1$ broadens (see Fig. 2(b)).
The ferromagnetism is absent. Near
$n=0.5$ we find peaks in the charge structure factor at $(\pi,\pi)$, which indicates a 
charge-ordered (CO) state.
Simultaneous peaks are found in the spin structure
factors at $(0,\pi)$ and $(\pi,0)$ pointing towards the existence of a nontrivial state with
simultaneous existence of charge- and spin-ordering.
All the results presented in this paper are for $N=32^2$, the stability of these results has been 
checked for system sizes up to $N=40^2$.

To further analyze the nature of electronic states we compute the total density of states (DOS) as,
\begin{eqnarray*}
N(\omega)=N^{-1}\sum_{i} \delta(\omega - \epsilon_i) \approx N^{-1}\sum_{i} \frac{\gamma/\pi}{\gamma^2 + (\omega-\epsilon_i)^2}.
\end{eqnarray*}
Here, $\epsilon_i$ denote the eigenenergies corresponding to the minimum energy configuration.
The $\delta$-function is approximated by a Lorentzian with width $\gamma$.
We use $\gamma = 0.04$ in the calculations.
A clean gap in the DOS is observed only for n=1 in the absence of $\lambda$ (see Fig. 2(c)).
In the FM regime,
a two-peak structure represents a shifted spin-up and spin-down band, 
which is consistent with the Stoner picture of
magnetism in Hubbard model. Eventually at low-density the DOS begins to resemble the free electron tight-binding
DOS. More interesting features are observed in the DOS at $\lambda=2$ shown in fig. 2(d).
The clean gap originating from the AFM state survives down to $n \sim 0.85$. The gap opens up once again at
quarter-filling ($n=0.5$). This correlates perfectly with the signatures found in the structure factor calculations shown in Fig. 2(b).

\begin{figure}[t]
\centerline{
\includegraphics[width=8.4cm , clip=true]{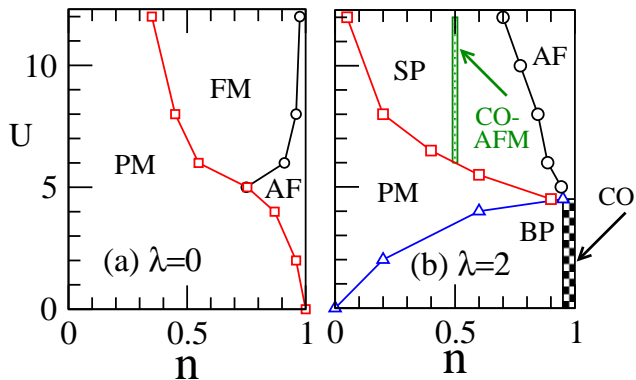}
}
\vspace{.1cm}
\caption{(Color online)~ The $U$-$n$ phase diagrams for the Hubbard-Holstein model
for (a) $\lambda=0$, and (b) $\lambda=2$. Ferromagnetic (FM), antiferromagnetic (AF),
paramagnetic (PM) phases are present in case of the pure Hubbard model ($\lambda=0$).
Single polaron (SP), bipolaron (BP) and charge ordered antiferromagnetic phases also become
stable for $\lambda=2$.
}
\end{figure}

The results for various $U$ at $\lambda=0$ and $\lambda=2$ are summarized into two phase diagrams.
The $U$-$n$ phase diagram for $\lambda=0$ is shown in Fig 3(a).
Antiferromagnetic, ferromagnetic
and paramagnetic states are found to be stable in agreement
with previous results on the Hubbard model in two-dimensions \cite{hirsch, stoner,Hubb-quarter}.
Fig. 3(b) shows the $U$-$n$ phase diagram for $\lambda=2$. For low $U$, the system becomes a
bipolaronic insulator. Although the $\lambda_c^{BP} \sim 2$ for a single BP, it can be much
lower for finite density, suggesting that it is easier to trap many bipolarons as compared
to a single BP \cite{sekhar-sk-pm}.
The charge ordered state at half-filling can be simply viewed as a checkerboard
arrangement of bipolarons, although the concept of an isolated BP does not really hold for
such large densities.
The charge-ordered state exists even below the critical coupling required for BP formation
due to the nesting feature
present in the Fermi surface at half filling.
The half-filled CO state undergoes a transition to an antiferromagnetic state near $U = 4$.
The region of antiferromagnetism grows with increasing $U$ in contrast to the pure Hubbard model.
The PM state still
exists for small $\lambda$ but the FM state is absent.
A large region of phase 
space is taken by the single polaronic state for large $U$.
No magnetism is found at low densities, since
these single polarons are magnetically non-interacting due to the large inter-polaronic separations.
At large densities, however, There are antiferromagnetic correlations between these single polarons.
This is consistent with the effective magnetic interactions found between two
single polarons (see Fig. 1(b)). These effective antiferromagnetic interactions are the origin of the
growth in the AFM regime near $n=1$.

\subsection{Half- and quarter-filling}

The Hubbard-Holstein model at and near half-filling has been studied 
previously \cite{muramatsu,half-filled,HH}.
The existence of spin- and charge-density waves was reported. The possbility for
an intermediate metallic phase was also reported in a one-dimensional model with dynamical effects for
lattice \cite{clay}.
Fig. 4(a) shows a $U-\lambda$ phase diagram at half-filling. The system is either charge ordered or
antiferromagnetic and therefore, the DOS is always gapped.
The boundary
separating the CO and the AFM states fits very well as U=$\lambda^{2}$, which happens to be the
boundary separating the SP and BP regime in the low density limit (see Fig. 1(a)). This
suggests that the CO phase can be viewed as a checkerboard pattern of bipolarons, at least for large
values of $\lambda$.
The origin of the CO or the AFM phase at small values of $U$ and $\lambda$ is related to the
existence of nesting in the Fermi surface with a nesting wavevector $ {\bf q} = (\pi,\pi)$.

\begin{figure}[t]
\centerline{
\includegraphics[width=8.8cm , clip=true]{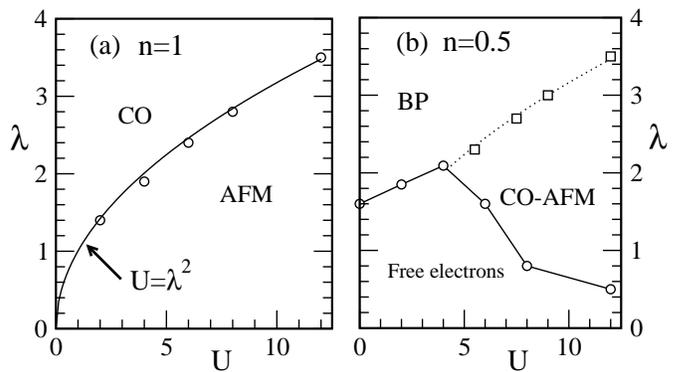}
}
\vspace{.1cm}
\caption{~ $\lambda-U$ phase diagrams at (a) half filling, and (b) quarter filling.
}
\end{figure}

The most interesting result of this paper is the observation of a CO-AFM state at quarter filling.
We plot the $U-\lambda$ phase diagram for quarter-filled system in Fig. 4(b).
Unlike the
half-filled case, the small U, small $\lambda$ regime corresponds to free electron behavior. The SP 
state is found to exist for large U and there is a large window where a charge-ordered AFM state exists.
We find that a self-consistent solution corresponding to a CO-FM state can exist only for $U>10$, 
but it is still higher in energy than the CO-AFM.
For $U<10$, the CO-FM state is not stable and therefore, the charge ordering occures only when it
is accompanied by an AFM ordering. This leads to a very interesting implication
for the effects of external magnetic field. Enforcing
ferromagnetism by applying an external magnetic field to the CO-AFM state would lead to a
melting of the charge order and
hence to a large negative magnetoresistence.

\begin{figure}[t]
\centerline{
\includegraphics[width=8.6cm , clip=true]{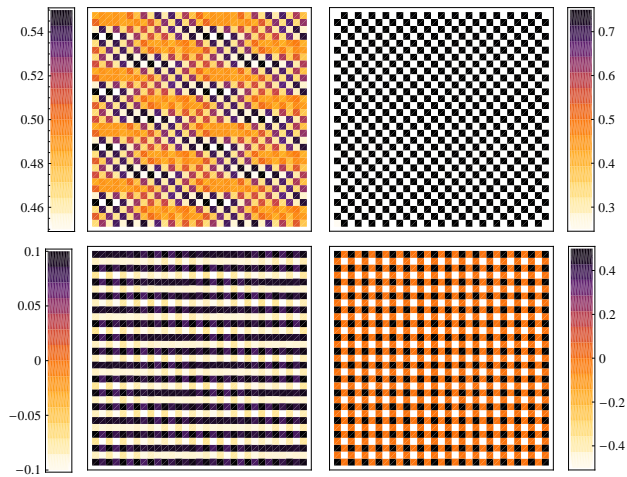}
}
\vspace{.1cm}
\caption{(Color online) ~ Real space patterns for charge density (upper row) and spin density 
(lower row) at $U=6$ and $\lambda = 1$ (left column),
and $\lambda = 2$ (right column). Note that the spin state for $\lambda=2$ is a G-type antiferromagnet
if one rotates the lattice by $45^{\circ}$ and consider the square lattice of "occupied" sites.
}
\end{figure}

To further investigate this unusual state at quarter-filling, we show the
real-space data for charge density $n_i$ and spin density $s_i$ in Fig. 5.
A weak charge-ordering is already present at $\lambda =1$ (see upper-left panel).
This charge ordering is accompanied
by a stripe-like spin ordering, where spins are arranged ferromagnetically along one direction and
antiferromagnetically along the other (lower-left panel). A very clear CO pattern emerges
for the larger value of $\lambda$ (upper-right panel), which occures together with an AFM arrangement of
the spins (lower-right panel). In a strong coupling scenario, it is easy to understand the CO-AFM
state within the picture of effective magnetic interaction between single polarons
presented in Fig. 1(b).
Assuming that the "occupied" sites in the charge ordered state can be viewed as single polarons, an effective antiferromagnetic interaction between them is strongest at distance $\sqrt{2}$, hence leading to a
magnetic structure which is a G-type AFM order for the square lattice constructed out of the "occupied" sites only.
For smaller values of
$\lambda$ the charge disproportionation in the CO state is much smaller and the effective magnetic interaction picture can not be pushed to this weak coupling regime.

\section{Conclusions}

We have presented groundstate properties of the Hubbard-Holstein model in two dimensions
in the adiabatic limit for the lattice distortions. We use a self consistent method for
generating the static lattice configurations in combination with a Hartree-Fock decoupling
of the Hubbard term.
Interestingly, the charge-ordered antiferromagnetic state that we find at quarter filling
was shown to be the groundstate for the LAO/STO interface in recent DFT
calculations \cite{zhicheng}.
Within our analysis the charge ordering in this state occures only in combination with the
AFM ordering, as we find that the CO-FM state is unstable. Therefore, the charge ordering could be
melted by applying an external magnetic field, leading to a large negative magnetoresistence.
We argue that this model is relevant for the LAO/STO interface since, (i) it provides a possibility
for the formation of magnetic moments, (ii) leads to a CO-AFM groundstate in agreement with the recent
LDA+U studies, and (iii) contains the possibility for large negative magnetoresistence via a
magnetic-field-induced melting of the charge ordered state.

\vspace{0.4cm}
\begin{center}
{\normalsize {\bf ACKNOWLEDGMENTS}}
\end{center}
We gratefully acknowledge useful discussions with Z. Zhong, G. Brocks and P. J. Kelly.
This work was financially supported by ``NanoNed'', a nanotechnology
programme of the Dutch Ministry of Economic Affairs and by the
``Nederlandse Organisatie voor Wetenschappelijk Onderzoek (NWO)''
and the ``Stichting voor Fundamenteel Onderzoek der Materie (FOM)''.

{}

\end{document}